\documentclass[twocolumn,twoside]{revtex4}
\usepackage{graphicx}
\usepackage{fancyhdr}
\begin{document}

\title{Charmoniumlike States at BESIII}

%

\author{W.~Imoehl\\
on behalf of the BESIII Collaboration}
\affiliation{Indiana University, Bloomington, IN, USA, 47401}

\begin{abstract}
  Despite being extensively studied since their initial discovery in 2003, the XYZ states do not yet have a clear interpretation. BESIII recently collected additional data above 4 GeV which can be used to further study these exotic hadron candidates. We present recent studies of $X(3872)\to\pi^0\chi_{c0}$, $e^+e^-\to K^+K^-J/\psi$, $e^+e^-\to$ light hadrons, $e^+e^-\to D^{*+}D^{(*)-}$, and on the $Z_{cs}(3985)$. 
\end{abstract}

\maketitle

\thispagestyle{fancy}


\newcommand{\RDOUBLE}{$\frac{\mathcal{B}(X(3872)\rightarrow\pi^0\chi_{c0})}{\mathcal{B}(X(3872)\to\pi^0\chi_{c1})}$}

\section{Introduction}

The XYZ states in the charmonium region have properties that make them inconsistent with the predicted charmonium spectrum. The first of these exotic hadron candidates to be discovered was the $X(3872)$ \cite{belle}. Its quantum numbers have been measured to be $J^{PC}=1^{++}$ \cite{lhcbNum}, but it has several properties that make it unlikely to be the $\chi_{c1}(2P)$. Its mass is below the predicted value for the $\chi_{c1}(2P)$, and its width has been measured to be $\Gamma=1.39\pm0.24\pm0.10$ MeV \cite{lhcbWidth}, which is narrower than expected for a pure charmonium state above $D\bar{D}$ threshold. In addition, the isospin violating decays $X(3872)\to\rho^0 J/\psi$ \cite{besRho} and $X(3872)\to\pi^0\chi_{c1}$ \cite{besPi} occur at a rate that is much larger than expected for a pure charmonium state. The measured mass of the $X(3872)$ is consistent with $D^{*0}\bar{D}^0$ threshold within experimental uncertainties, which has prompted predictions that the $X(3872)$ is a molecular meson, a compact tetraquark, the $\chi_{c1}(2P)$, or some mixture of these three scenarios \cite{xyzRev}.

The vector meson $Y$ states overpopulate the predicted distribution for the charmonium system. Since the vector mesons have $J^{PC}=1^{--}$, they can be produced directly in $e^+e^-$ collisions, so their masses and widths can be determined by fitting the measured $e^+e^-$ cross section values. BESIII has used this method to resolve the single $Y(4260)$ resonance into two resonances, the $Y(4230)$ and the $Y(4360)$, using the measured $e^+e^-\to\pi^+\pi^-J/\psi$ cross section values \cite{besY}. The theoretical explanations for these states include compact tetraquarks, molecular mesons, and hybrid mesons \cite{xyzRev}.

The $Z_c(3900)^\pm$ and $Z_c(3900)^0$ states were observed at BESIII in the processes $e^+e^-\to\pi\pi J/\psi$ in the invariant $\pi J/\psi$ spectrum \cite{zc,zc0}. Since the $Z_c(3900)$ is an isovector, it clearly cannot be described as a pure charmonium state, and must have at least four constituent quarks. The leading interpretations for the $Z_c$ states are molecular mesons or compact tetraquarks \cite{xyzRev}.

To study the quark configuration of the XYZ states, we use data collected by the BESIII detector, which is located at the Beijing Electron Positron Collider (BEPCII). The BESIII detector records symmetric $e^+e^-$ collisions and covers 93\% of the $4\pi$ solid angle for photons and charged particles. This is an excellent environment for studying XYZ states because they are produced nearly at rest with relatively small background levels. This enables BESIII to reconstruct complicated decay modes of these states. In the remainder of this paper, we discuss the recent searches for new $X(3872)$ and $Y(4230)$ decay modes at BESIII. We also report the recent discovery of the $Z_{cs}(3985)^+$, the strange partner to the $Z_c(3900)$. 

\section{Search for $X(3872)\to\pi^0\chi_{c0}$}

\begin{figure*}
  \includegraphics[scale=1.0]{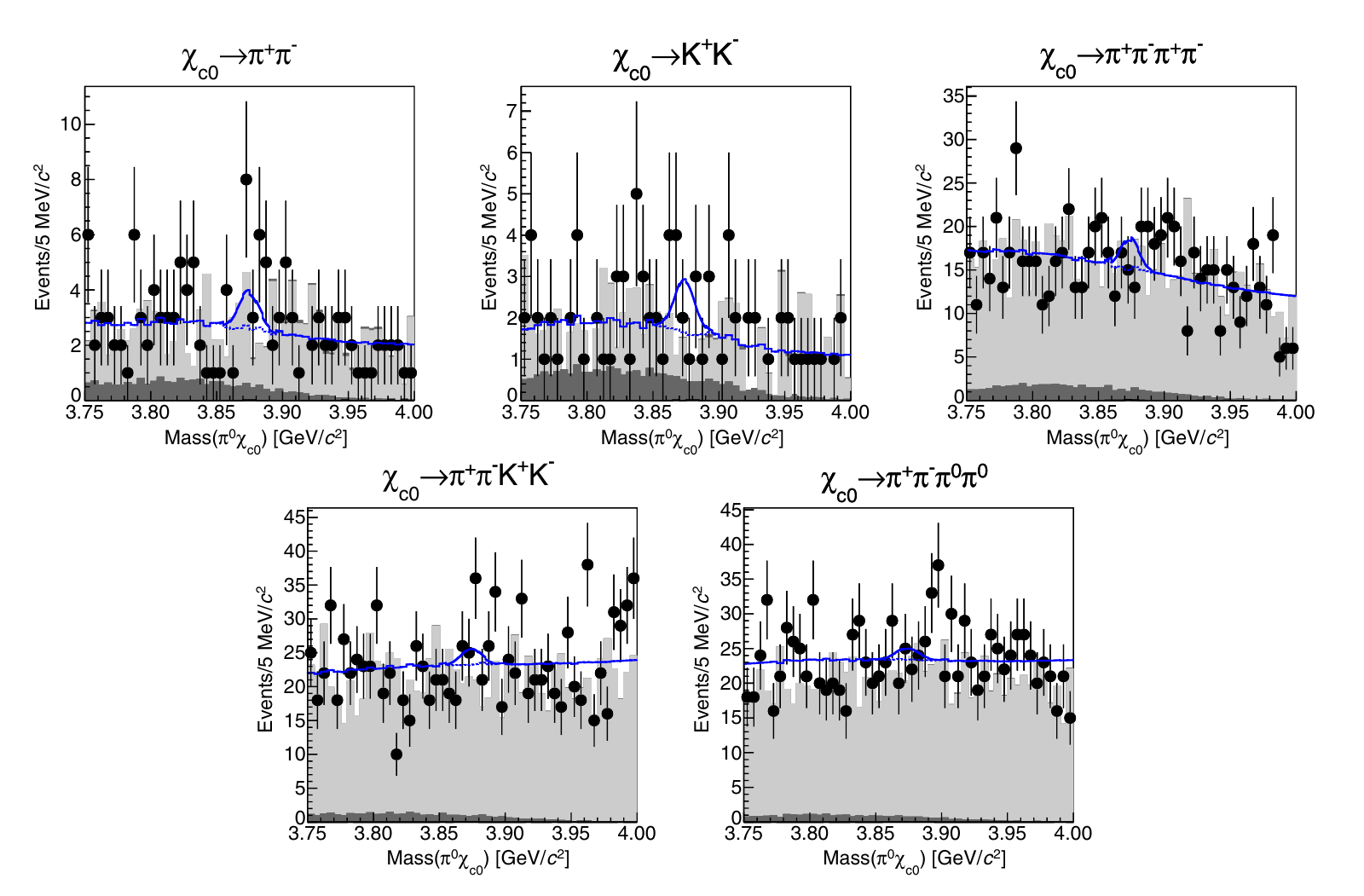}
  \caption{Fit to the $\pi^0\chi_{c0}$ invariant mass distribution for the search of $X(3872)\to\pi^0\chi_{c0}$. The solid line is the fit with a signal component while the dashed line is a fit with just the background component. No significant signals are found.}
  \label{fig:pi0chic}
\end{figure*}

The decay $X(3872)\to\pi^0\chi_{c0}$ is predicted to be sensitive to the quark configuration of the $X(3872)$, since if the $X(3872)$ has four constituent quarks, the ratio of branching fractions is predicted to be \RDOUBLE$\approx 3$, while for pure $c\bar{c}$, this decay should be forbidden. Using the production process $e^+e^-\to\gamma X(3872)$, this analysis searches for the decay $X(3872)\to\pi^0\chi_{c0}$, where the $\chi_{c0}$ decays hadronically to five final states. No significant signals are observed, so we place a 90\% confidence level upper limit of \RDOUBLE$<4.5$ using the fit results in Fig. \ref{fig:pi0chic}. This is too large to rule out any interpretation of the $X(3872)$, but it is the most sensitive search for this decay mode to date \cite{mine}. We also perform the first searches for the decays $X(3872)\to\pi^+\pi^-\chi_{c0}$ and $X(3872)\to\pi^0\pi^0\chi_{c0}$ but find no significant signals. 

\section{Observation of $Y(4230)$ in $e^+e^-\to K^+K^- J/\psi$}

\begin{figure}
  \includegraphics[scale=1.0]{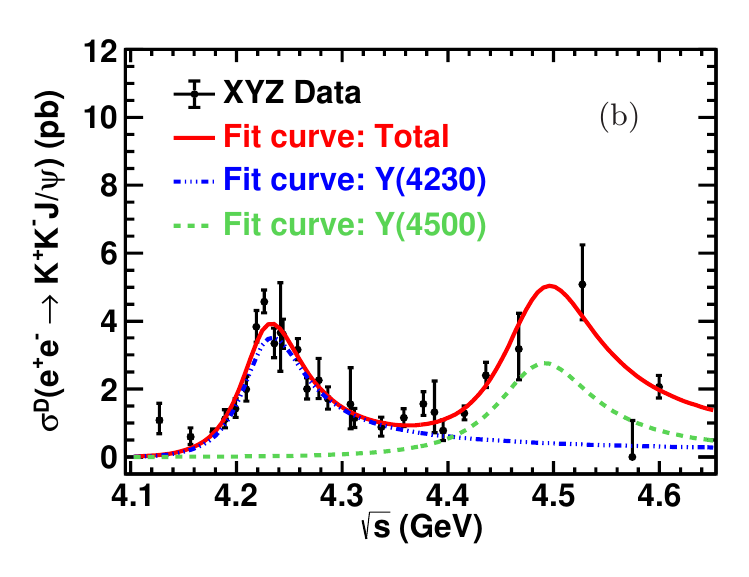}
  \caption{Measured cross section for $e^+e^-\to K^+K^- J/\psi$ showing the total fit (red line) and the components due to the $Y(4230)$ (dashed blue) and $Y(4500)$ (dashed green).}
  \label{fig:kkjpsi}
\end{figure}

To probe the strange quark content of the $Y(4230)$, we measure the cross section of $e^+e^-\to K^+K^-J/\psi$ to search for resonant contributions. In addition to the $Y(4230)$, this analysis could be sensitive to predictions from conventional charmonium models with 5S-4D mixing, hadronic molecule models, and tetraquark models that all predict a state near 4.5 GeV$/c^2$. The measured cross section values are shown in Fig. \ref{fig:kkjpsi}, where we observe a clear $Y(4230)$ peak \cite{kkjpsi}. The cross section clearly rises after the $Y(4230)$ peak, so this rise is fit with a Breit-Wigner. The mass from the Breit-Wigner is consistent with a peak near 4.5 GeV$/c^2$, but more data is needed to draw firm conclusions about what is happening in this higher $E_{\rm cm}$ region. 

\section{Search for Charmoniumlike States Decaying to Light Hadrons}

\begin{figure*}
  \includegraphics[scale=1.0]{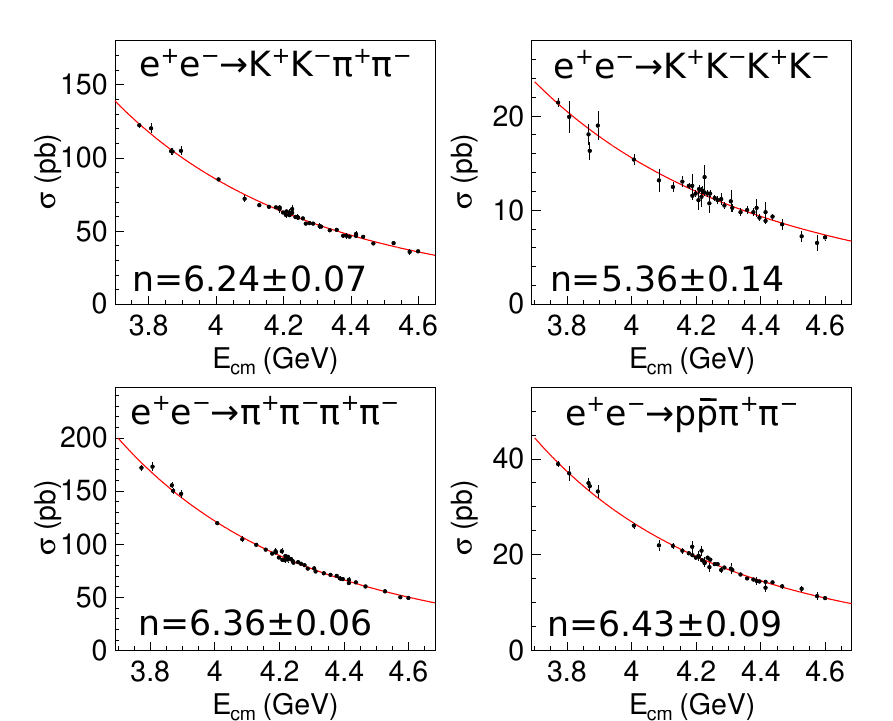}\includegraphics[scale=1.0]{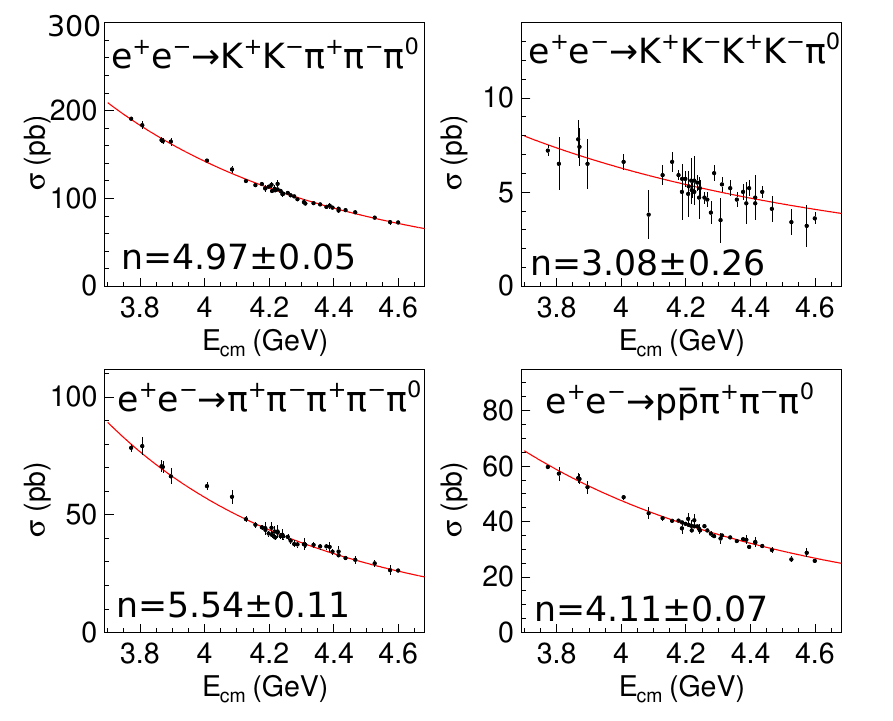}
  \caption{Measured cross section values (points) for 8 light hadron final states fit with $\sigma_{cont}^{expected}$ (red line). We do not observe any resonant structures for any of the measured cross sections.}
  \label{fig:light}
\end{figure*}

No light hadron decays have been found for any charmonium or charmoniumlike states above 4 GeV. The purpose of this analysis is to search for light hadron decays above 4 GeV to probe the light quark content of the $Y$ states. To do this, we measure the cross sections for eight light hadron final states. The measured values for each cross section are fit with $\sigma_{cont}^{expected}=|A_{cont}|^2$ where 
\begin{equation}
  A_{cont}=\sqrt{\frac{f_{cont}}{(E_{\rm cm}/4.226\textrm{ GeV})^n}}
\end{equation}
where $f_{cont}$ and $n$ are floating parameters in the fit and $E_{\rm cm}$ is the measured center-of-mass energy. We do not observe any resonant structures in these energy regions \cite{light}, as shown in Fig. \ref{fig:light}.

\section{Measurement of $\sigma(e^+e^-\to D^{*+}D^{(*)-})$}

\begin{figure}
  \includegraphics[scale=1.0,trim={6.5cm 0cm 0cm 0.06cm},clip]{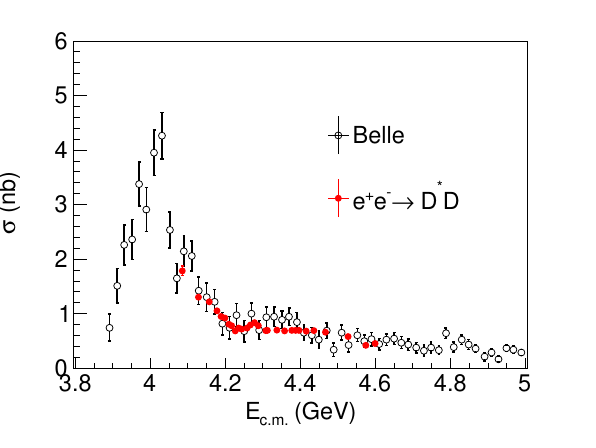}\\\includegraphics[scale=1.0,trim={6.5cm 0cm 0cm 0cm},clip]{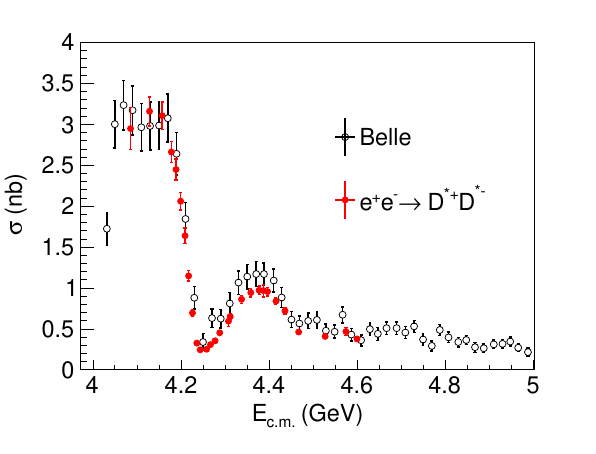}
  \caption{Measured open charm cross section values from Belle (black points) and BESIII (red points). The recent BESIII measurements have a good overall agreement with the Belle results, but with much smaller error bars. The error bars shown in the plot are the quadrature sum of the statistical and systematic uncertainties.}
  \label{fig:dd}
\end{figure}

The conventional vector meson charmonium states above open charm threshold predominantly decay to open charm final states. By contrast, the vector meson charmoniumlike states have large decay rates to hidden charm final states. Open charm cross section measurements can be used in a coupled channel analysis to determine the pole positions of resonances in these energy regions. BESIII recently measured more precise cross section values for $e^+e^-\to D^{*+}D^{-}$ and $e^+e^-\to D^{*+}D^{*-}$ \cite{dd}, as shown in Fig. \ref{fig:dd}. The cross section measurements are consistent with previous Belle measurements \cite{belleD} but have smaller error bars, which will improve the sensitivity of future coupled channel analyses of the charmonium system.

\section{The Charged and Neutral $Z_{cs}(3985)$ at BESIII}

\begin{figure}
   \includegraphics[scale=0.8]{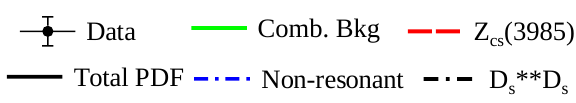}\\ 
   \includegraphics[scale=0.8]{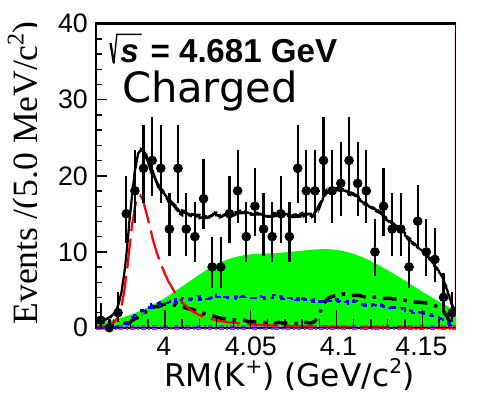}\includegraphics[scale=0.85]{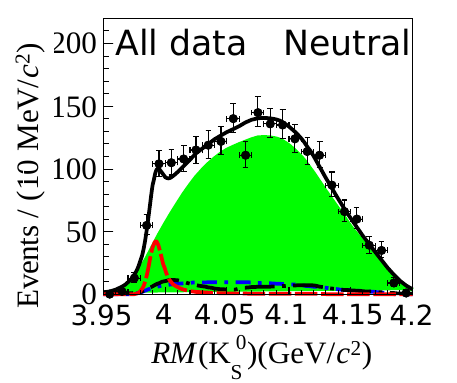}
   \caption{Measurements of $Z_{cs}(3985)^\pm$ (left) and $Z_{cs}(3985)^0$ (right). The charged structure has a significance of $5.3\sigma$, while the neutral case has a significance of $4.6\sigma$.}
   \label{fig:zcs}
\end{figure}

BESIII recently reported a $5.3\sigma$ observation of the $Z_{cs}(3985)^\pm$ in $e^+e^-\to K^+(D_s^-D^{*0}+D_s^{*-}D^0)$ \cite{zcs}, as shown in Fig. \ref{fig:zcs} left. This is a candidate to be the strange partner of the $Z_c$ states, and would have a minimal quark content of $c\bar{c}s\bar{u}$. Additionally, BESIII has searched for the neutral partner to the $Z_{cs}^\pm$ in $e^+e^-\to K_s^0(D_s^-D^{*+}+D_s^{*-}D^+)$, and found $4.6\sigma$ evidence for the $Z_{cs}(3985)^0$ \cite{zcs0}, as shown in Fig. \ref{fig:zcs} right. The minimal quark content of the neutral state is $c\bar{c}s\bar{d}$. The measured mass and width for the charged and neutral $Z_{cs}$ candidates are shown in Table \ref{tab:zcs}. These are consistent with theoretical predictions that the neutral $Z_{cs}$ should have a higher mass than the charged $Z_{cs}$.

\begin{table}
  \caption{Measured mass and width values for each of the $Z_{cs}$ candidates}
\begin{tabular}{|c|c|c|}\hline
  State & Mass [MeV/$c^{2}$] & Width [MeV]\\\hline
  $Z_{cs}(3985)^+$ & $3985.2^{+2.1}_{-2.0}\pm 1.7$ & 13.8$^{+8.1}_{-5.2}\pm4.9$ \\\hline
  $Z_{cs}(3985)^0$ & $3992.2\pm1.7\pm1.6$ & $7.7^{+4.1}_{-3.8}\pm 4.3$\\\hline
\end{tabular}
\label{tab:zcs}
\end{table}

\section{Summary}

BESIII continues to be active in studying the XYZ states in the charmonium system. We have recently searched for new decays of both the $X(3872)$ and the $Y(4230)$. No evidence was found for the decays $X(3872)\to\pi^0\chi_{c0}$, $X(3872)\to\pi\pi\chi_{c0}$, or $Y(4230)\to$ light hadrons, but the process $Y(4230)\to K^+K^-J/\psi$ was observed for the first time. The open charm cross sections $e^+e^-\to D^{*+}D^{-}$ and $e^+e^-\to D^{*+}D^{*-}$ have been measured with a higher precision than the previous Belle results, which will improve the sensitivity to the XYZ states in future coupled channel analyses. We recently observed the $Z_{cs}(3985)^\pm$ and found evidence for its neutral partner. An accelerator upgrade is planned for 2024 that will increase the luminosity by up to a factor of 3 and give access to energies up to 5.6 GeV. This will give the experiment access to more charmed baryons and the capability to search for new $Y$ states at higher energies.

\bigskip 

\end{document}